# Benchmarking Universal Interatomic Potentials on Zeolite Structures

Shusuke Ito[1], Koki Muraoka[1, *], Akira Nakayama[1, *]

[1]*Department of Chemical System Engineering, The University of Tokyo, Tokyo 113-8656, Japan*

## Abstract

Interatomic potentials (IPs) with wide elemental coverage and high accuracy are powerful tools for high-throughput materials discovery. While the past few years witnessed the development of multiple new universal IPs that cover wide ranges of the periodic table, their applicability to target chemical systems should be carefully investigated. We benchmark several universal IPs using equilibrium zeolite structures as testbeds. We select a diverse set of universal IPs encompassing two major categories: (i) universal analytic IPs, including GFN-FF, UFF, and Dreiding; (ii) pretrained universal machine learning IPs (MLIPs), comprising CHGNet, ORB-v3, MatterSim, eSEN-30M-OAM, PFP-v7, and EquiformerV2-lE4-lF100-S2EFS-OC22. We compare them with established tailor-made IPs, SLC, ClayFF, and BSFF using experimental data and density functional theory (DFT) calculations with dispersion correction as the reference. The tested zeolite structures comprise pure silica frameworks and aluminosilicates containing copper species, potassium, and organic cations. We found that GFN-FF is the best among the tested universal analytic IPs, but it does not achieve satisfactory accuracy for highly strained silica rings and aluminosilicate systems. All MLIPs can well reproduce experimental or DFT-level geometries and energetics. Among the universal MLIPs, the eSEN-30M-OAM model shows the most consistent performance across all zeolite structures studied. These findings show that the modern pretrained universal MLIPs are practical tools in zeolite screening workflows involving various compositions.

## Introduction

Driven by advances in computational resources and extensive software development, atomistic computational approaches are used in nearly all disciplines of materials science[1]. Among various methods for calculating potential energy surfaces, particularly popular is density functional theory (DFT) using Perdew–Burke–Ernzerhof (PBE) functional[2,3]. It has been chosen in several material databases, such as Materials Projects[4], AFLOW[5], OQMD[6], Alexandria[7], and OMat24[8], owing to its good performance with relatively low computational cost. To reduce the noncovalent interaction error and static correlation error, recently developed SCAN-based functionals[9–11], such as $r^2$SCAN[10], are emerging as alternatives for next-generation high-throughput computations[12–16].

However, DFT calculations require a high computational cost for modeling large structures with more than a few hundred atoms[17,18]. In such cases, less computationally expensive empirical interatomic potentials (IPs) have been a natural alternative methodology.

The advantage of empirical IPs in terms of speed stems from their simpler form of expression and better scaling. Most of the empirical IPs are designed for a certain class of chemical systems. A drawback of those tailor-made IPs is the inability to describe diverse chemical environments[19]. For example, bond formation and cleavage in most cases fall outside their applicable domain[20]. While some IPs emphasize the transferability[21,22], it is nontrivial to reuse a parameter set devised for a certain chemical system in completely different systems. For these reasons, applying tailor-made IPs to diverse systems in an unbiased manner requires careful consideration.

Universal IPs are designed to describe a wide range of chemical environments containing various elements on the periodic table by employing a single set of parameters. The Universal Force Field (UFF)[23] and the Extensible Systematic Force Field[24] were early developmental efforts aimed at achieving this objective. While comprehensive electronic structure calculations were used as reference data, they have not succeeded in achieving high accuracy across the entire periodic table[25,26], likely due to the simplicity of their functional forms.

Recently, a new wave of IP development aiming to cover the whole periodic table has emerged in the realm of analytic IPs[19,27] and machine learning IPs (MLIPs)[28–30]. GFN-FF, a recently developed universal analytic IP, integrates semiempirical quantum-mechanical methods with empirical covalent bond terms. Its parameters are fitted to DFT results of about 8000 structures to handle as much as 86 elements[19].

Preferred Potential (PFP) is one of the first successful MLIPs covering the arbitrary combination of the most elements in the periodic table. Its graph neural network architecture[31] has been trained against a diverse dataset generated by DFT to capture complex topological and electronic interactions[28]. Another MLIP, CHGNet, predicts magnetic moments from atomic coordinates and species, incorporates the inferred charge information into atomic features, and uses them to improve potential predictions[29]. Other pretrained universal MLIPs, MatterSim, Orb, eSEN-30M-OAM, and UMA were consecutively reported[8,30,32–34]. The goal of these universal MLIPs is to enable calculations for any chemical system with high accuracy and low computational cost without further training. Given their reported applications for electrolytes[35], porous materials[36,37], and crystal structures[38], the universal MLIPs have huge potential to advance materials science. Because their general accuracy is still in discussion, some studies try to benchmark their applicability for broad structure datasets[39,40].

In this study, we benchmark some universal analytic IPs and universal MLIPs for the zeolite structures. Zeolites are porous aluminosilicate crystalline materials that play a central role in tackling several environmental problems. Atomistic simulations of zeolites typically involve more than hundreds of atoms, which makes the utilization of the analytic IPs[41,42] or MLIPs[43] a reasonable choice. Additionally, zeolites present an ideal testbed for this benchmarking due to their consistent composition coupled with a rich diversity of atomic environments involving organic and inorganic guest cation species. This enables a reasonable assessment of the performance of universal IPs, without the influence of compositional offsets. The tested structures involve pure silica zeolites, aluminosilicate zeolites containing potassium, organic cations, and copper species. The relaxed geometries and energies are compared with experimental data or DFT calculations. While universal analytic IPs do not provide reasonable structures or energetics in some cases, pretrained universal MLIPs show remarkable performance in mimicking the results of DFT calculations. This study confirms that some pretrained MLIPs can be used for high-throughput calculations of zeolites.

# Methods

## Dataset

Experimental structural data and thermochemical data were obtained from the previous papers[42,44]. Pure silica zeolite structures were obtained from the International Zeolite Association[45]. To construct the copper ion-exchanged CHA zeolites, we choose $Cu^{2+}$ and $[Cu(OH)]^+$ as the cation species. First, we place the $Cu^{2+}$ between pairs of Al atoms; then, we position $[Cu(OH)]^+$ near isolated or remaining Al atoms. We generate as many placement patterns for the copper species as possible when defining the Al pairs in the structure, while minimizing the number of $[Cu(OH)]^+$ ions. The structure data of ERI zeolites containing potassium and OSDA were generated following a previous paper[46].

## Computational details

All zeolite structures are relaxed using the MPRelax setting in pymatgen[47] version 2025.5.2, utilizing PBE exchange-correlation functional[2] within the projector augmented-wave (PAW) framework[48,49] as implemented in the Vienna Ab initio Simulation Package (VASP) 6.2.1[50–53] with DFT-D3 correction[54,55]. The energy cutoff is set to 520 eV, and the convergence criterion for the energy is set to $5 \times 10^{-5}$ eV/atom. Calculation errors are automatically managed using custodian[47] version 2025.5.12. For example, the tetrahedron method is initially applied for Brillouin zone integration, and in cases where it fails to determine the Fermi level, Gaussian smearing is used instead. The parameters described above are the same as those used in the early version of the Materials Project[56].

Geometry and cell parameter optimization using analytic IPs is performed using the GULP software[57,58] without applying symmetry constraints. A strain-based optimization is employed, utilizing a Newton-Raphson optimizer with BFGS Hessian updates. The optimizer is configured to switch to the RFO method when the gradient norm is less than 0.1. All convergence threshold values are set to their default values.

Universal MLIPs are called through the Python packages CHGNet, orb-models, MatterSim, and fairchem, corresponding respectively to the use of CHGNet[29], ORB-v3[33], MatterSim[32], and the models from OMat24/OC22[8,59]. Structural optimizations are performed using the FIRE algorithm with a FrechetCellFilter implemented in ASE[60] until the maximum residual force on atoms falls below 0.02 eV/Å.

# Results and Discussion

## Reproducibility of experimental results for pure silica zeolites

Using zeolites as a testbed, this study benchmarks IPs encompassing three major categories: (i) universal analytic IPs, including GFN-FF, UFF, and Dreiding; (ii) pretrained universal MLIPs, comprising CHGNet, ORB-v3, MatterSim, eSEN-30M-OAM (hereafter called eSEN), PFP-v7, and EquiformerV2-lE4-lF100-S2EFS-OC22 (hereafter called EqV2); (iii) tailor-made IPs, including SLC, ClayFF, and BSFF. Table 1 classifies the IPs that can be applied to zeolites. The table also includes tailor-made MLIPs for zeolites[20,61], while we do not cover them in this study. As detailed in the Supplementary Information, the parameters of DFT calculations used to generate training data for pretrained universal MLIPs are largely consistent, facilitating direct comparison.

We select silica as our initial test set due to its role as the fundamental building block of zeolites. Despite sharing the same $SiO_2$ composition, silica zeolites display remarkable structural diversity in bond angles and bond lengths depending on the polymorphs. Accurately predicting these structural characteristics is the first requirement for reliable potentials for zeolites. We obtain experimental structure data for eight pure silica structures from a previous study[42] (see Supplementary Table 1). The corresponding crystal structures are relaxed using DFT, tailor-made IPs, universal analytic IPs, and universal MLIPs (Fig. 1).

We use D3 correction[54] in DFT and MLIP calculations because it has been argued that DFT includes systematic errors without van der Waals corrections in zeolites[62].

**Table 1. Classification of the potentials used in zeolites**

|  | Tailor-made | Universal |
|---|---|---|
| DFT |  | PBE, SCAN* |
| MLIP | Pure silica zeolites [61*], water-loaded acidic zeolites [20*] | PFP, CHGNet, MatterSim, ORB, eSEN-30M-OAM, EquiformerV2-lE4-lF100-S2EFS-OC22, UMA* |
| Analytic IP | SLC, ClayFF, BSFF | UFF, Dreiding, GFN-FF |

*: Not covered in this study

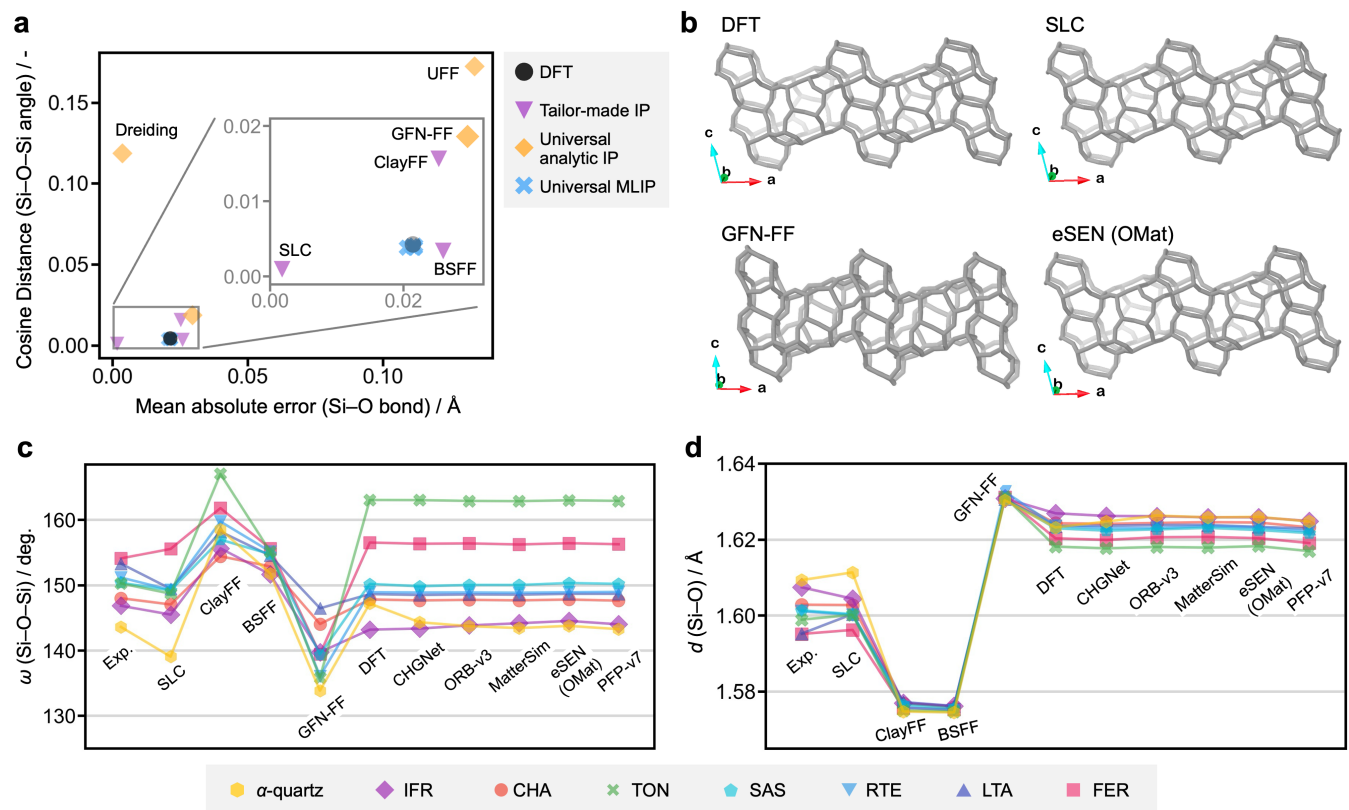

**Fig. 1. Geometric comparison between experimental and calculated data.** (a) The mean absolute error in Si–O bond lengths and the cosine distance in Si–O–Si angles are used as metrics. Black circles represent a value of DFT calculation using PBE functional with D3 correction, while purple triangles, orange diamonds, and blue crosses denote results from tailor-made IPs, universal analytic IPs, and universal MLIPs, respectively. (b) RTE-type zeolite structures relaxed by DFT with PBE+D3, SLC, GFN-FF, and eSEN with D3 correction. Distribution of (c) average bond angles and (d) bond lengths in pure silica zeolites, calculated using several potentials excluding the UFF and Dreiding. Only structures with available experimental data are included. Each data point represents the average value for a unique zeolite topology, assigned a distinct color.

Fig. 1a visualizes the mean absolute errors and cosine distances of various potentials against experimental data. Among all tested potentials, SLC potential[63–66] achieves structures closest to the experimental data. This is reasonable as SLC is explicitly fitted with the experimental data of one of the silica polymorphs ($\alpha$-quartz)[63]. The other tailor-made IPs, ClayFF[67] and BSFF[68], showed still good but less consistent structures compared to SLC. Although ClayFF is one of the tailor-made IPs extensively employed for zeolitic structures[69–71], it has been reported to produce the unrealistic Si–O–Si angles for some zeolites[42]. As the tested universal MLIPs are trained on DFT with PBE functional, they are scattered in a close area to each other and to the DFT result (Fig. 1a).

Among the universal analytic IPs, UFF[23] and Dreiding[72] show significant deviations from the experimental results. GFN-FF provides better agreement with the experimental data, showing comparable performance to that of ClayFF. GFN-FF, however, causes distortions in some structures, including RTE-type zeolite (Fig. 1b), leading to a change in symmetry. In contrast, DFT with PBE+D3, SLC, and eSEN produce reasonable RTE-type zeolite structures as shown in Fig. 1b. According to a structure matching algorithm implemented in pymatgen, the relaxed RTE-type zeolite structures obtained using SLC, PBE+D3, and eSEN are considered equivalent, whereas the structure obtained with GFN-FF is not. Because all the tested universal MLIPs are trained on DFT results, they provide results similar to DFT, as shown in Fig. 1a, c, and d. Both universal MLIPs and DFT show acceptable agreement with experimental data, considering that the experiments were performed at finite temperature.

Fig. 1c and d provide the distributions of average bond angles and bond lengths of pure silica zeolites. Each data point signifies the average value of a zeolite structure. SLC well reproduces both Si–O bond lengths and Si–O–Si bond angles, in agreement with the result in Fig. 1a. Other tailor-made IPs, ClayFF and BSFF, show values different from the experimental data. The range of Si–O–Si angles obtained with BSFF matches the experimental values more closely compared to ClayFF. This improvement is likely due to the additional bond angle terms incorporated into BSFF[68] to better capture the structural characteristics of zeolites. Although GFN-FF was close to ClayFF in the averaged errors (Fig. 1a), the respective values seem to be very different, as shown in Fig. 1c and d: GFN-FF underestimates the Si–O–Si angles and overestimates the Si–O lengths, while ClayFF shows the opposite trend.

While DFT with PBE+D3 seems to reproduce experimental data of Si–O–Si bond angles relatively well, the distribution of Si–O bond length is somewhat different from experimental data. Again, this discrepancy can arise from the experimental error and the differences in temperature. All tested MLIPs+D3 show the same trend as the DFT+D3, suggesting that pretraining is successful in reproducing DFT results for the silica structures.

Next, we focus on the energetics. Among silica polymorphs, $\alpha$-quartz is recognized as the thermodynamically most stable phase, while silica zeolites are metastable at standard conditions. The relative energy, referenced to $\alpha$-quartz, serves as a reliable metric for assessing the accuracy of various potentials against experimental data[42,68].

We calculate the relative energies of some silica zeolites with available experimental thermodynamic data[44] (see Supplementary Table 1) using different analytic IPs, MLIPs, and DFT calculations (Fig. 2 and Table 2). As in previous studies, we ignore the vibrational contribution and directly compare the DFT relative energies with the experimental ones[42,74].

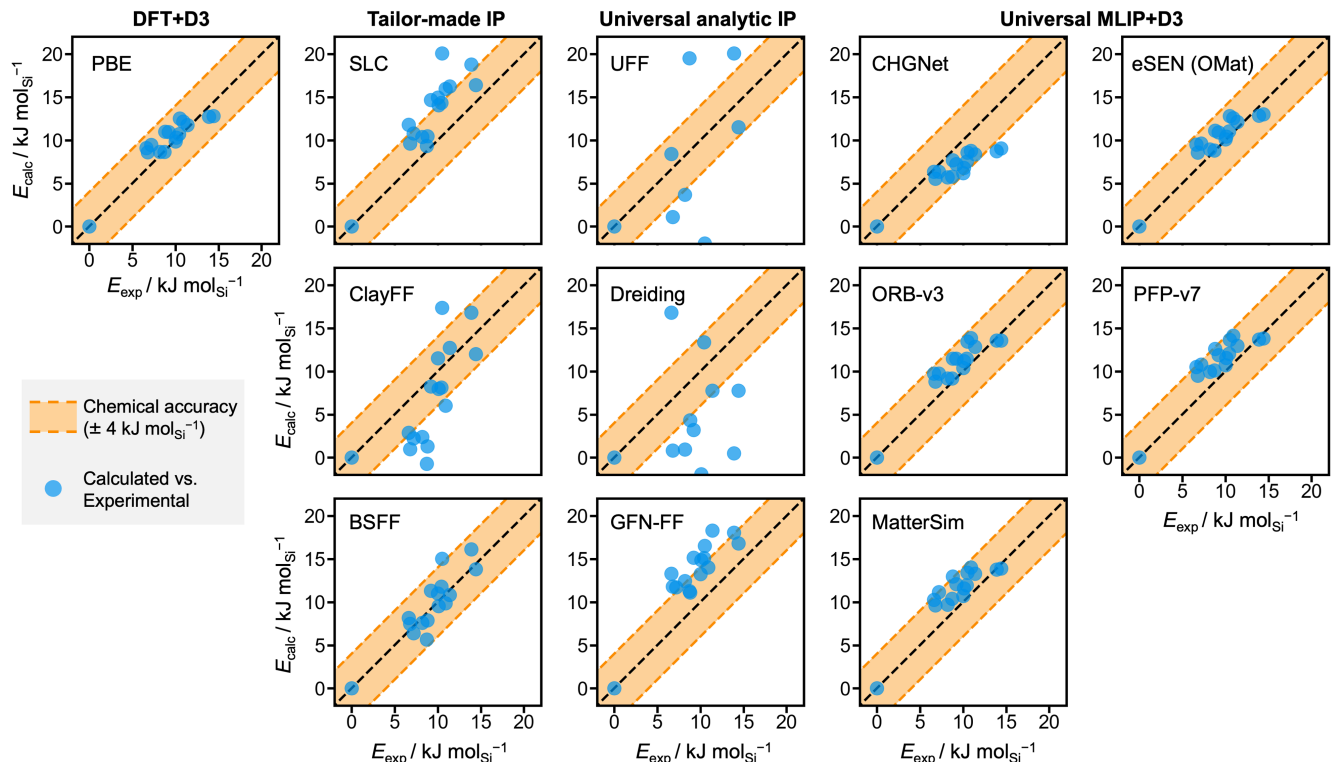

**Fig. 2. Comparison of calculated and experimental values of relative energies of zeolites.** Scatter plots show $E_{calc}$ vs. $E_{exp}$ for several calculation results, DFT with PBE+D3, tailor-made IP, universal analytic IPs, and universal MLIPs with D3 (universal MLIP+D3). The orange region indicates chemical accuracy (±4 kJ mol$^{-1}$)[73]. The y–y plots present a limited range for clarity in comparison. All data points are shown for IPs except UFF and Dreiding, for which only points within the plotted range are displayed.

**Table 2. Root-mean-squared errors (RMSE) of relative energies against the experimental data.** The IP with the smallest RMSE within each category is highlighted in bold.

|  | DFT+D3 | Tailor-made IP | | | Universal analytic IP | | | Universal MLIP+D3 | | | | |
| --- | --- | --- | --- | --- | --- | --- | --- | --- | --- | --- | --- | --- |
|  | PBE | SLC | ClayFF | BSFF | UFF | Dreiding | GFN-FF | CHGNet | ORB-v3 | MatterSim | eSEN (OMat) | PFP-v7 |
| RMSE / kJ mol$_{Si}^{-1}$ | **1.43** | 4.41 | 4.68 | **1.75** | 15.99 | 15.36 | **4.53** | 2.84 | 1.88 | 2.44 | **1.55** | 2.38 |

DFT using the PBE functional with D3 correction shows the highest accuracy in reproducing experimental relative energies. One of the universal MLIPs, eSEN, exhibits the second-smallest RMSE value (Table 2). As shown in Fig. 2, all universal MLIPs behave very close to the DFT, likely due to the successful training of the universal MLIPs using DFT results.

All tailor-made IPs show good agreement with experimental data. BSFF outperforms the other tailor-made IPs in terms of energetics, as reported in a previous study[68]. GFN-FF gives the most accurate predictions among the universal analytic IPs. In contrast, UFF and Dreiding are unable to provide reliable predictions of relative energies, as evidenced by both the large RMSE values and the lower correlation in the y–y plots (see Table 2, Fig. 2, and Supplementary Fig. 1).

For pure silica zeolites, the tailor-made SLC potential reproduces experimental bond lengths and angles. DFT with PBE+D3 best predicts experimental relative energies, and with slightly smaller error by eSEN, a universal MLIP. Given their computational cost, using DFT instead of SLC or universal MLIPs for modeling of pure silica zeolites is generally unnecessary unless there is a specific purpose that justifies it, as noted in a previous study on the SLC potential[42].

## Reproducibility of DFT results for pure silica zeolites

To expand the test coverage to zeolite structures without experimental data, hereafter, we employ the relative energy calculated by PBE+D3 as the reference data to assess the accuracy of the IPs. We use structures with pure silica composition for almost all zeolite topologies, which were obtained from the International Zeolite Association database[45]. Fig. 3 and Table 3 reveal that all tailor-made IPs show systematic errors, especially in relatively unstable zeolites. Part of the reason would be that the tailor-made IPs have been constructed without relatively unstable zeolite structures, and their parameters have been fitted to experimental data observed at finite temperature, while DFT results describe the ground-state properties at 0 K.

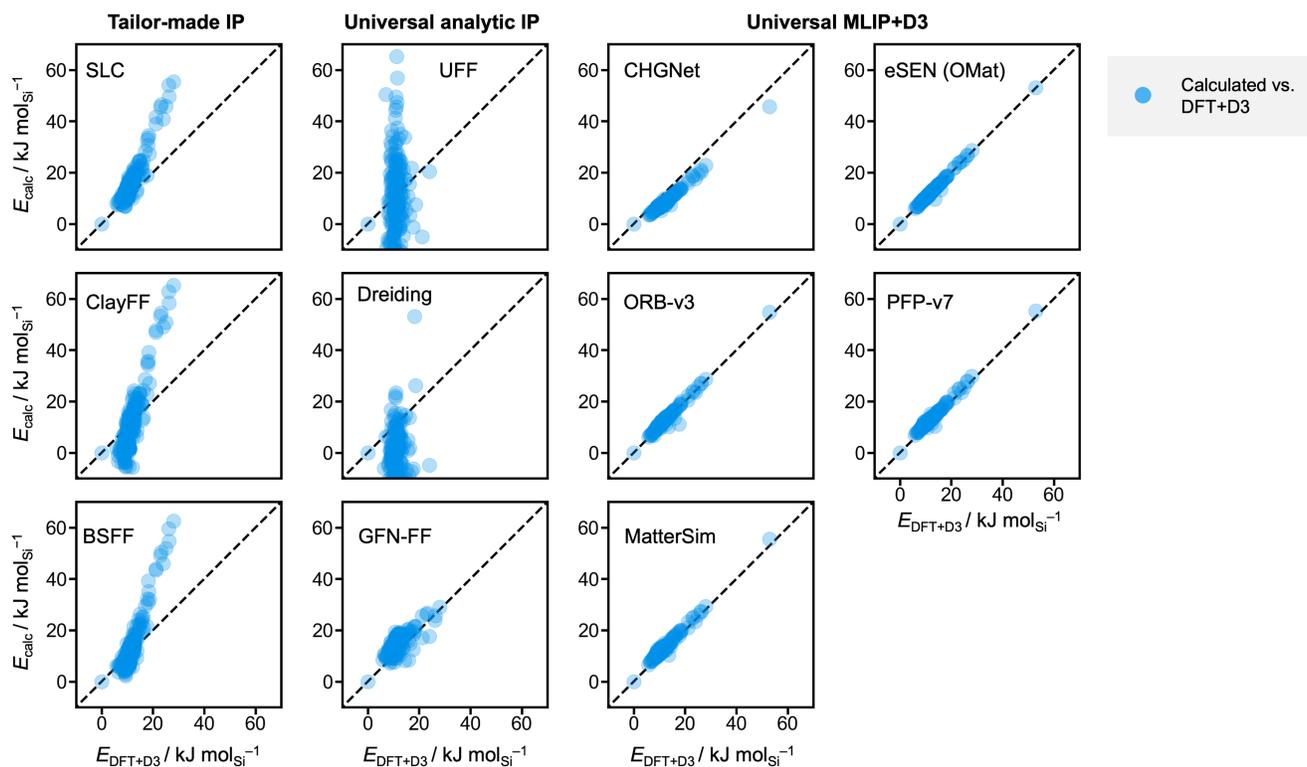

**Fig. 3. Comparison of relative energies calculated by DFT with PBE+D3 and other IPs.** The y–y plots present a limited range for clarity in comparison. Some of the data points for GFN-FF, UFF, and Dreiding are outside of the plot (see Supplementary Fig. 2 for the complete data).

The y–y plots of UFF and Dreiding show very little correlation with the DFT results, indicating that they have difficulty capturing the thermodynamic properties of zeolites (see Supplementary Fig. 2). GFN-FF is able to reproduce the stability of zeolite structures well for most cases, but it shows a remarkable discrepancy for SOS-type and RWY-type zeolites (see Supplementary Fig. 3a). Their relaxed structures using GFN-FF are distorted, as shown in Supplementary Fig. 3b. It seems that GFN-FF struggles to handle structures with steep Si–O–Si angles, especially those in three-membered rings. These results show that the broad application of the universal analytical IPs needs care.

**Table 3. RMSE of potentials against DFT.** The IP with the smallest RMSE within each category is highlighted in bold.

| | Tailor-made IP | | | Universal analytic IP | | | Universal MLIP+D3 | | | | |
|---|---|---|---|---|---|---|---|---|---|---|---|
| | SLC | ClayFF | BSFF | UFF | Dreiding | GFN-FF | CHGNet | ORB-v3 | MatterSim | eSEN (OMat) | PFP-v7 |
| RMSE / kJ mol$_{Si}^{-1}$ | 7.40 | 8.98 | **7.60** | 44.55 | 40.86 | **24.44** | 3.60 | 1.01 | 1.49 | **0.44** | 1.33 |

For the calculation of pure silica zeolites, eSEN shows the best performance among all tested universal IPs and MLIPs. CHGNet seems to entail a systematic error, as shown in Fig. 3. We attribute this error to the lower prediction accuracy of the absolute energy value for relatively stable topologies (see Supplementary Fig. 4). It is of note that DFT reference results always include deviations from experimental data, as described in Fig. 1 and a previous study[42].

## Reproducibility of DFT results for guests containing zeolites

So far, we have shown that universal MLIPs are capable of handling silica zeolite frameworks well. Because a key advantage of universal potentials is their ability to treat a wide range of chemical systems, not only in terms of various conformations but also diverse chemical compositions, we next focus on structures with more chemical diversity. One important application for atomistic modeling of zeolites is the modeling of zeolite catalysts. As an example of zeolite catalysts, we construct 347 copper-introduced CHA-type zeolite structures. $Cu^{2+}$ and $[Cu(OH)]^+$ species are randomly introduced in aluminosilicate CHA-type zeolites with different Al distributions. Another important application of atomistic modeling for zeolites is the modeling of zeolites with as-synthesized structures[75–81]. We generate 1,190 ERI-type zeolite structures containing potassium and hexane-1,6-bis(trimethylazanium) as organic structure-directing agent (OSDA), obtained from previous research[46]. The Cu/CHA and K-OSDA/ERI structures have the same chemical compositions but differ in their aluminum distributions and/or the location of guest cation species. We relax all structures using DFT with PBE+D3 and perform single-point calculations using GFN-FF and several universal MLIPs. Unlike in previous discussions, here we include an MLIP model from OpenCatalyst (OC), as cell optimization is not required. To evaluate their performance, we calculate relative energies for each type of zeolitic structure. Reference phases are defined as the most stable structures in DFT results (Fig. 4).

Although GFN-FF shows good prediction accuracy in CHA and ERI pure silica zeolites, it exhibits little correlation in the y-y plots and large RMSE in Cu/CHA and K-OSDA/ERI zeolites (Supplementary Fig. 5). It should be noted that GFN-FF has been constructed using DFT with B97-3c functional as reference data[19,82], whereas the other universal MLIPs were trained on a dataset using DFT with PBE functional. Thus, the discrepancy between GFN-FF and DFT results may be partly attributed to the difference in the reference data.

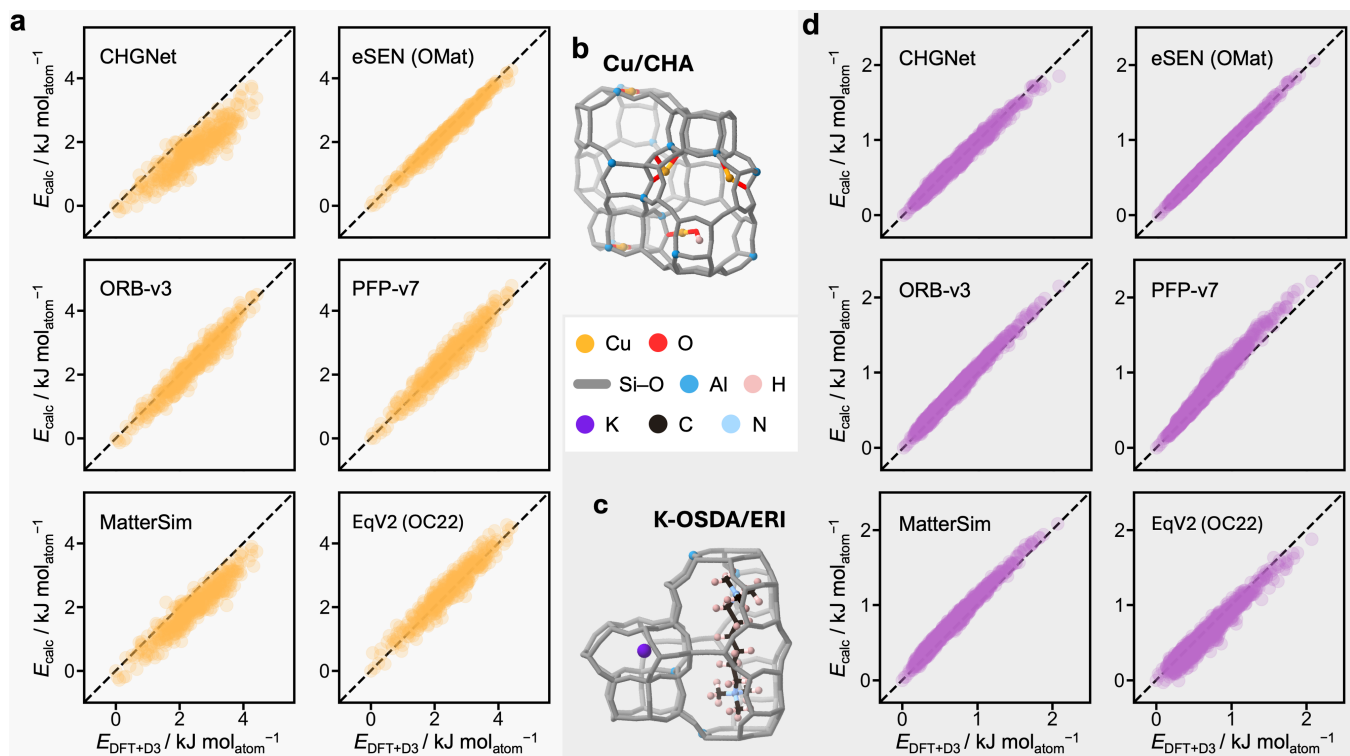

**Fig. 4. Comparison of relative energies, with respect to the most stable structures calculated by DFT.** (a) y-y plots for 347 copper ion-exchanged CHA zeolites and (b) the representative structure. (d) y-y plots for 1,190 potassium and organic structure-directing agents (OSDAs)-containing ERI zeolites and (c) the representative structure.

**Table 4. RMSE of MLIPs with respect to DFT.** The MLIP with the smallest RMSE is highlighted in bold.

|  |  | CHGNet | ORB-v3 | MatterSim | eSEN(OMat) | PFP-v7 | EqV2(OC22) |
|---|---|---|---|---|---|---|---|
| RMSE / kJ mol$_{atom}^{-1}$ | Cu/CHA | 0.76 | 0.24 | 0.52 | **0.14** | 0.24 | 0.26 |
|  | K-OSDA/ERI | 0.04 | 0.07 | 0.09 | **0.02** | 0.11 | 0.09 |

Fig. 4 shows the performance of universal MLIPs for guest cation-containing zeolites. While all MLIP models reproduce the DFT results well, Table 4 illustrates that the eSEN model again excels all the other MLIPs in both Cu/CHA and K-OSDA/ERI structures. When comparing the performance of Cu/CHA and K-OSDA/ERI, it appears to be more difficult to accurately predict the DFT results of Cu/CHA zeolites, as shown in Fig. 4 and Table 4. We assume that it is a more complicated task to calculate the structures with transition metal species than those with alkali metal or organic cations.

While we do not consider the diversity of chemical bonds in our scope, it is insightful that the eSEN model from OMat24, one of the universal MLIPs, outperforms other universal IPs and achieves high accuracy at the DFT level for pure silica, Cu/CHA, and K-OSDA/ERI zeolites. As the development of universal MLIPs continues to accelerate, new models with better performance will likely be proposed. Since our guest-containing structures are unlikely to be used in training data for present and future universal MLIPs, they can also be utilized for evaluating future potentials, avoiding data leakage.

# Conclusion

This study provides a comprehensive benchmark of several universal IPs for zeolite materials ranging from pure silica frameworks to guest cation-containing structures. Through a comparative analysis of structural accuracy and relative energetics, several key insights for their utilization are revealed.

Among tailor-made IPs, the SLC potential continues to be the best option for high speed and good accuracy for reproducing the experimental data of pure silica zeolites, whereas ClayFF and BSFF hold systematic deviations in local structures, as observed in some distorted topologies.

Universal IPs show divergent behavior. Rule-based approaches such as UFF and Dreiding break down for both geometry and thermodynamics, underscoring the difficulty of parameterizing a single analytic potential across the whole periodic table. GFN-FF, which incorporates semi-empirical quantum-mechanical terms, improves substantially on this baseline and captures many frameworks energetics reliably, yet still struggles with highly strained rings and guest cations-containing zeolites.

On the other hand, universal MLIPs are consistently closer to the experimental and DFT data. Among them, the eSEN model from OMat24 data delivers the best reproduction of both geometries and relative energies across all tested zeolite structures. Other modern MLIP models—ORB-v3, PFP-v7, MatterSim, CHGNet, and EqV2(OC22)—also reproduce the DFT results with acceptable degrees of error. It is worth noting that the training data for universal MLIPs, as well as part of the reference data in our benchmark, are based on DFT calculations, which themselves involve intrinsic errors[42]. As DFT methodologies continue to advance, MLIPs trained on a more accurate level of theory are expected to achieve accuracy even closer to experimental observations.

Altogether, these findings suggest a practical hierarchy for zeolite modelling. For silica frameworks, SLC remains the most efficient option. For more complicated systems, state-of-the-art universal MLIPs serve as practical tools to enable high-throughput materials exploration.

# Data availability

The dataset of all structures relaxed by tested IPs and their calculated energies is publicly available on Zenodo (https://doi.org/10.5281/zenodo.17075635).

# Acknowledgments

This work is supported by JSPS KAKENHI (22K14751), JST PRESTO (JPMJPR2378), and JST SPRING (JPMJSP2108).

# Author Information

## Corresponding Authors

**Koki Muraoka** − *Department of Chemical System Engineering, Graduate School of Engineering, The University of Tokyo, Tokyo 113-8656, Japan*; orcid.org/0000-0003-1830-7978 Email: muraok_k@chemsys.t.u-tokyo.ac.jp

**Akira Nakayama** − *Department of Chemical System Engineering, Graduate School of Engineering, The University of Tokyo, Tokyo 113-8656, Japan*; orcid.org/0000-0002-7330-0317; Phone: +81 (0) 358417270; Email: nakayama@chemsys.t.u-tokyo.ac.jp

## Author

**Shusuke Ito** − *Department of Chemical System Engineering, Graduate School of Engineering, The University of Tokyo, Tokyo 113-8656, Japan*

# Author contribution

S.I. conducted the investigation. S.I. handled the visualization of the results. K.M. and A.N. supervised the project. The original draft was written by S.I. and K.M., with review and editing contributions from S.I., K.M., and A.N.

# Additional information

## Supplementary Information

Summary of the details of several analytic IPs and the parameters of DFT calculations used to generate training data for pretrained universal MLIPs, along with additional supplementary tables and figures.

# Supplementary Information for

# Benchmarking Universal Interatomic Potentials on Zeolite Structures

Shusuke Ito[1], Koki Muraoka[1, *], Akira Nakayama[1, *]

[1]*Department of Chemical System Engineering, The University of Tokyo, Tokyo 113-8656, Japan*

## Details of Interatomic Potentials (IPs)

The parameters in SLC potential were fitted empirically to structural and physical properties of $\alpha$-quartz[1], $\alpha$-Al$_2$O$_3$[2], and micas[3] with auxiliary parameters specific to the Si$^{4+}$–O$^{1.4-}$ and Al$^{3+}$–O$^{1.4-}$ interactions constructed for $\alpha$-quartz and sillimanite[4]. ClayFF was parameterized for clay minerals, including aluminosilicates and those are chemically similar to zeolites[5]. BSFF uses the parameters of atomic interaction from ClayFF with the introduction of additional O–Si–O and Si–O–Si bending terms in order to improve the reproducibility for $\alpha$-quartz and zeolite structures[6].

Dreiding was developed for organic, biological, and simple inorganic molecules, using a minimal set of parameters based on idealized hybridization geometries and covalent radii[7]. UFF was designed to cover the entire periodic table, with parameters derived from theoretical atomic properties such as effective radii and ionization energies[8]. The parameters of both IPs are constructed without fitting to experimental or DFT results. GFN-FF has the parameters fitted to reproduce the DFT (B97-3c[9]) results of a versatile dataset containing about 8000 molecular structures, ranging from small molecules to large transition-metal complexes[10].

PFP was trained on a proprietary dataset generated from DFT calculations using parameters largely consistent with those of the Materials Project, except that a fixed Gaussian smearing was applied, whereas the Materials Project used either the Tetrahedron method or Gaussian smearing depending on the structures[11]. CHGNet was trained on the publicly available trajectories of the structure optimization (named MPtrj) generated through constructing the Materials Project[12]. Although some Orb-v3 models were trained on a dataset comprising materials from OMat24[13], MPtrj[12], and Alexandria[14], we used a model trained only on OMat24, following the recommendation[15]. It is worth noting that the parameters for the calculation to construct datasets of OMat24, MPtrj, and Alexandria were chosen to be compatible with the Materials Project[13–16], although those in the OMat24 include some important exceptions[13]. The model from OMat24 used in this study is eSEN-30M-OAM (hereafter called eSEN). This model is trained on OMat24 dataset, and fine-tuned by using the dataset from MPtrj and Alexandria[13]. MatterSim is trained on datasets from Materials Project, Alexandria, and newly generated structures, which are sampled through classical MD simulations and computed using DFT under conditions consistent with the Materials Project[17]. From the OC22, we employed the EquiformerV2-lE4-lF100-S2EFS-OC22 (hereafter called EqV2) model for only single-point energy calculations of copper-introduced and potassium and OSDA-containing zeolites. This model was trained exclusively on the OC22 dataset[18], which was generated using computational settings same with those in the Materials Project. Note that EqV2 was not used for relaxation of pure silicas because it does not support direct stress calculations. We believe that the consistency in training dataset construction enables a more reliable comparison with the other models used in this study.

Table 1. Zeolites used to compare with experimental data.

| Experimental structural data[19] | Experimental thermodynamic data[20] |
|---|---|
| α-quartz | α-quartz |
| CHA | CHA |
| FER | FER |
| IFR | IFR |
| LTA | AFI |
| RTE | CFI |
| SAS | EMT |
| TON | ISV |
|  | ITE |
|  | MEI |
|  | MEL |
|  | MWW |
|  | STT |
|  | MFI |
|  | MTW |

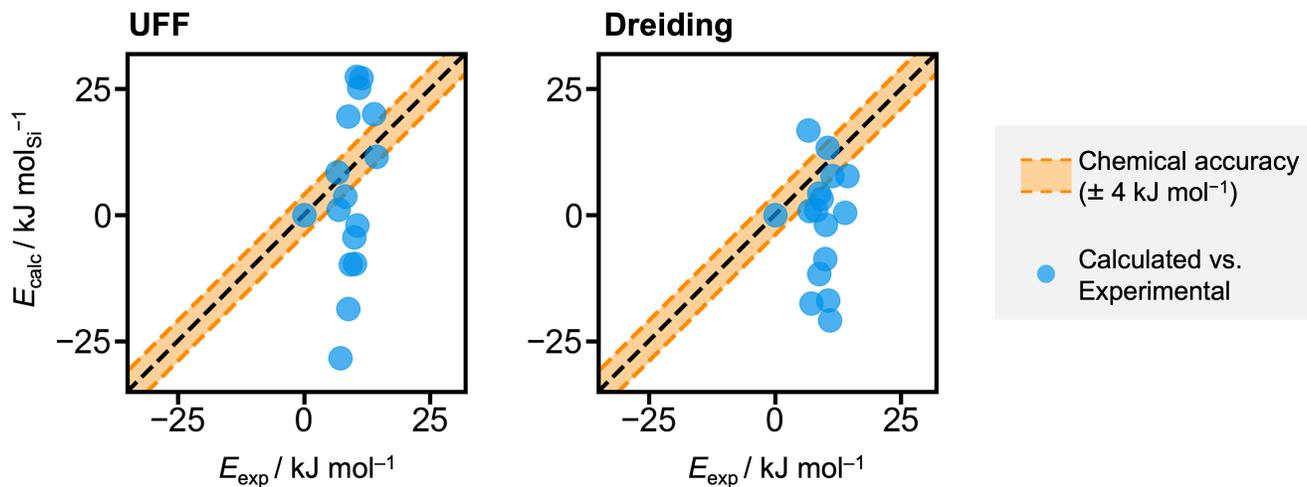

Fig. 1. Comparison of calculated and experimental values of relative energies of zeolites. Scatter plots show $E_{calc}$ vs. $E_{exp}$ of UFF and Dreiding. The y–y plot shown in Fig. 2 presents a limited range for clarity in comparison. In contrast, this visualizes the full data range.

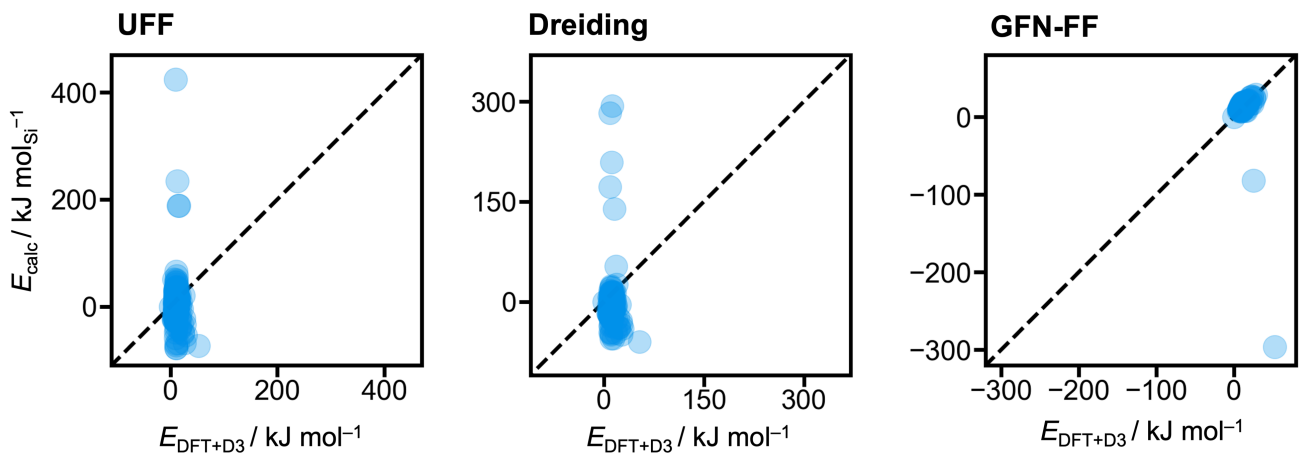

Fig. 2. Comparison of relative energies calculated by DFT of PBE functional with D3 correction and universal IPs (UFF, Dreiding, and GFN-FF). The y–y plot shown in Fig. 3 presents a limited range for clarity in comparison. In contrast, this visualizes the full data range.

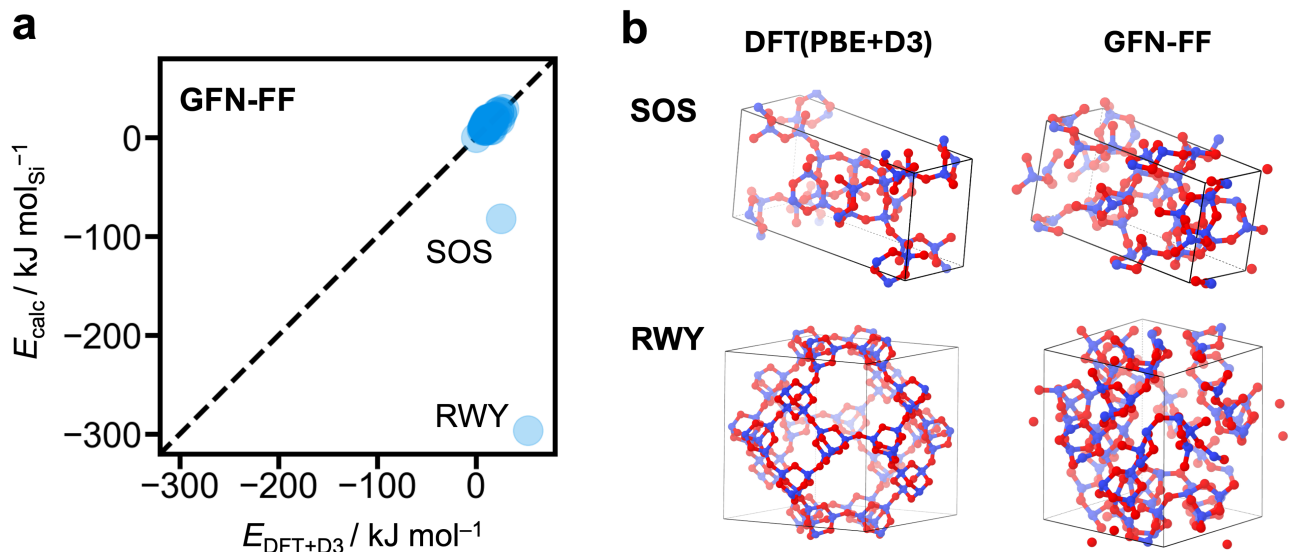

Fig. 3. Comparison of (a) relative energies of pure silica zeolites and (b) relaxed SOS and RWY zeolite structures obtained through DFT of PBE functional with D3 correction and GFN-FF.

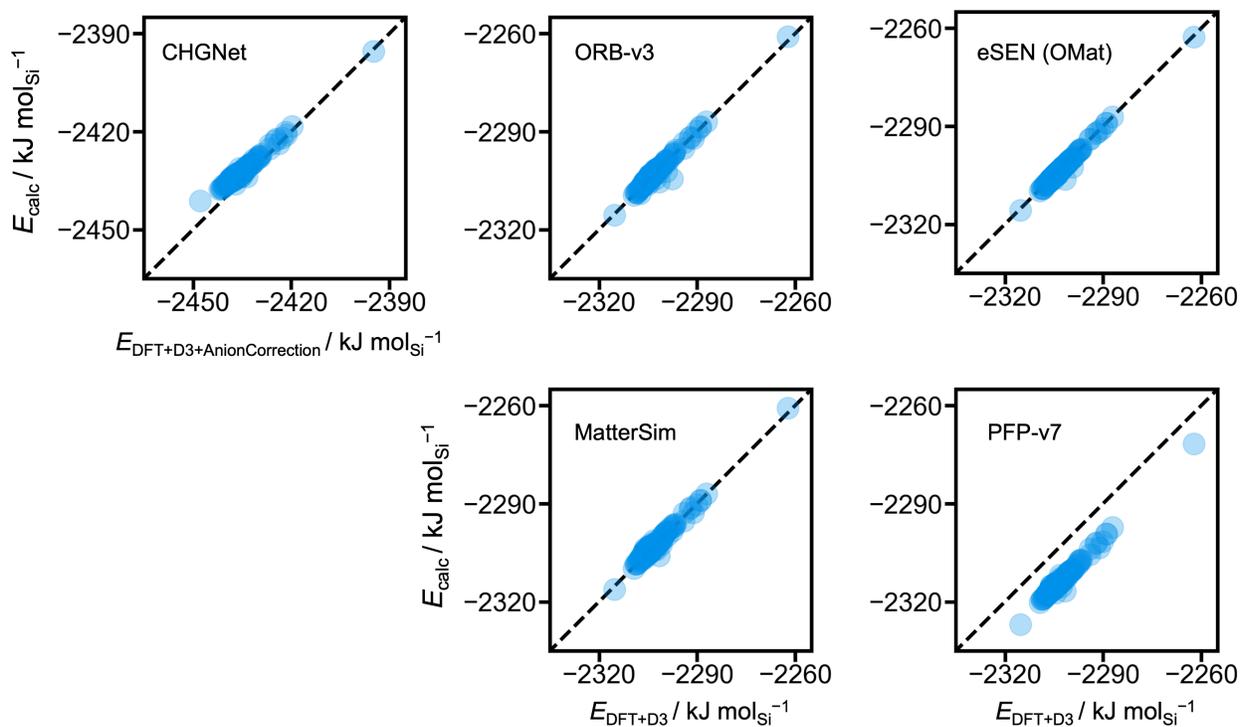

Fig. 4. Comparison of energies of pure silica zeolites calculated by DFT of PBE functional with D3 correction and universal MLIPs (CHGNet, ORB-v3, MatterSim, eSEN, and PFP-v7). Note that the horizontal axis of y-y plots for CHGNet uses Materials Project's anion correction, as CHGNet has been trained on corrected energies. CHGNet overestimates the energetics of relatively stable pure silica structures. The calculated value on PFP-v7 is corrected appropriately, as it uses isolated atoms as the energy reference.

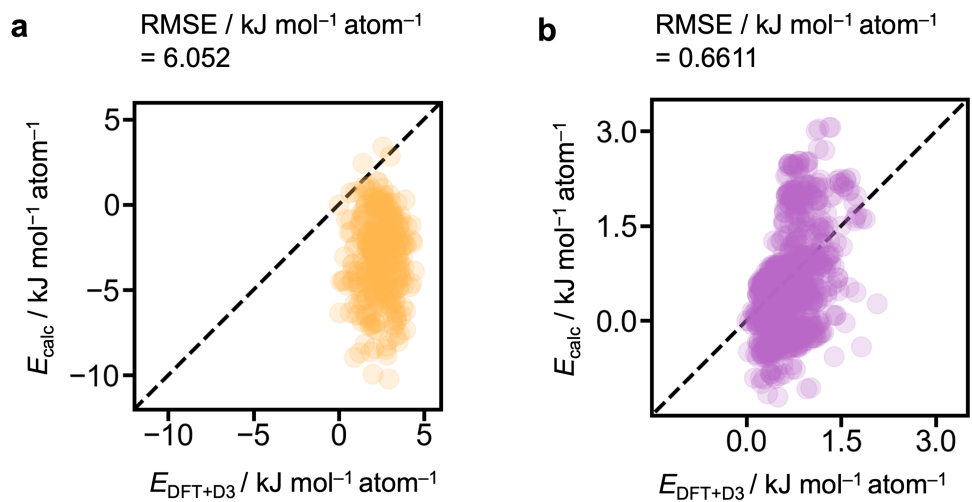

Fig. 5. Comparison of relative energies between DFT and GFN-FF, with respect to the most stable structures calculated by DFT in (a) 347 copper ion-exchanged CHA zeolites and (b) 1190 potassium and OSDAs-containing ERI zeolites.